\documentclass[aps,preprint,floatfix,%
showpacs,showkeys,amssymb]{revtex4}

\usepackage{graphicx}

\begin{document}

\date{\today} 

\def\ba{\begin{array}}
\def\ea{\end{array}}
\def\be{\begin{equation}\begin{array}{l}}
\def\ee{\end{array}\end{equation}}
\def\bea{\begin{equation}\begin{array}{l}}
\def\eea{\end{array}\end{equation}}
\def\f#1#2{\frac{\displaystyle #1}{\displaystyle #2}}
\def\om{\omega}
\def\omm{\omega^a_b}
\def\we{\wedge}
\def\de{\delta}
\def\De{\Delta}
\def\va{\varepsilon}
\def\omb{\bar{\omega}}
\def\la{\lambda}
\def\vv{\f{V}{\la^d}}
\def\si{\sigma}
\def\t{T_+}
\def\v{v_{cl}}
\def\m{m_{cl}}
\def\n{N_{cl}}
\def\bi{\bibitem}
\def\c{\cite}
\def\sa{\sigma_{\alpha}}
\def\ua{\uparrow}
\def\da{\downarrow}
\def\mua{\mu_{\alpha}}
\def\ga{\gamma_{\alpha}}
\def\g{\gamma}
\def\G{\Gamma}
\def\ora{\overrightarrow}
\def\pa{\partial}
\def\ov{\ora{v}}
\def\al{\alpha}
\def\bt{\beta}
\def\R{R_{eff}}
\def\th{\theta}
\def\na{\nabla}

\def\muu{\f{\mu}{ed}}
\def\E{\f{edE(\tau)}{\om}}
\def\t{\tau}

\baselineskip 6.2mm
\parskip 1.5mm

\title{Correlation functions for the $XY$ model in a Magnetic Field}

\author{Wei Zhang $^{1,2}$ and H. A. Fertig $^1$}

\affiliation{$^1$Department of Physics, Indiana University, Bloomington, IN 47405\\
$^2$Department of Physics and Astronomy, Ohio University, Athens, OH 45701}

\begin{abstract}
Recent studies of the two-dimensional, classical
$XY$ magnet in a magnetic field suggest that it has three 
distinct vortex phases: a linearly confined 
phase, a logarithmically confined phase, and a free vortex phase.
In this work 
we study spin-spin correlation functions in this model by analytical
analysis and numerical simulations to search for signatures
of the various phases. In all three phases,
the order parameter is nonzero and $<\cos(\th({\bf r}_1))\cos(\th({\bf r}_2))>$ remains
nonzero 
for $r \equiv |{\bf r}_1-{\bf r}_2|\rightarrow \infty$, indicating the expected long range order. 
The correlation function for transverse fluctuations of the spins,
$C(r)=<\sin(\th({\bf r}_1))\sin(\th({\bf r}_2))>$, falls exponentially in all three phases.
A renormalization group analysis suggests that the logarithmically confined phase
should have a spatially anisotropic correlation length.
In addition, there is a generic 
anisotropy in the prefactor which is always present.
We find that this prefactor
anisotropy becomes rather strong in the presence of a magnetic
field, masking the effects of any anisotropy in the correlation
length in the simulations.
\end{abstract}

\maketitle

\section{Introduction} The
two-dimensional classical XY model is
important because it serves as a paradigm for many
systems, including easy-plane magnets, two dimensional
solids, thin film superconductors and superfluids, 
magnetic bubble arrays, and certain one-dimensional quantum
systems.  Of particular interest in these systems is the behavior
of their topological excitations, which in the $XY$ model
are vortices.
In spite of the Mermin-Wagner-Hohenberg theorem \c{mw}, 
which forbids long-range order (i.e., a non-vanishing
order parameter) at any finite temperature
in two dimensions, a (Kosterlitz-Thouless, or KT) phase transition
occurs in this system, in which the vortices go from
a bound pair phase at low temperature 
to an unbound vortex phase at high temperature \c{kt}.
While the order parameter vanishes in both phases,
correlation functions show different behaviors in the different phases. 
In the high temperature phase,
the correlations decrease with increasing distance according to 
an exponential law \c{kt,corrxyh}, while in the low temperature
phase,  the correlations decrease 
according to a power law \c{kt,jose}. 

A non-vanishing order parameter can trivially be restored by
the application of a magnetic field which tends to order the
$XY$ spins.  The behavior of the vortices in this situation \c{jose}
has recently been revisited \c{herb1,herb2,herb3} via renormalization
group (RG) and simulation studies, leading to the conclusion
that vortices can still unbind due to thermal fluctuations in
this system, in a two-step process.  At the lowest
temperatures, vortex-antivortex pairs are linearly confined by a string of
overturned spins.  As temperature is increased, these strings first
undergo a proliferation transition, but the vortices remain
confined due to a residual logarithmic attraction.  With increasing
temperature this attraction is overcome and the vortices
deconfine.  Importantly, unlike the KT transition,
these are not phase transitions in the thermodynamic sense,
as they do not lead to singularities in the free energy \c{herb1,herb2}.
Nevertheless, the transitions can introduce singularities
in correlation functions \c{ruelle}, and lead to dramatically
different behaviors for the physical systems in which the
$XY$ model with a magnetic field is realized.  This includes
the behavior of 
bilayer thin film superconductors \c{wei}, the bilayer quantum Hall system
\c{gm}, and bosons in a linear optical lattice, tunnel-coupled
to a bulk superfluid reservoir \c{kingshuk}.

Since a standard way to characterize phases and phase transitions 
is via the behavior of correlation functions, it is natural to
search them for signatures of the deconfinement transitions.
In this paper, we study spin-spin correlation 
functions of the system in all three phases through
both analytical analysis and numerical simulations. 
Due to the external field, all the three phases will have a
non-zero order parameter, and for a symmetry-breaking
term (i.e., magnetic field coupling) of the form
$-h \cos{\theta({\bf r})}$, with $\theta({\bf r})$
the angular variable of the $XY$ spin located at
lattice site ${\bf r}$, one easily confirms that
the correlation function \break
$<\cos(\th({\bf r}_1))\cos(\th({\bf r}_2))>\rightarrow const.$, 
for $r=|{\bf r}_1-{\bf r}_2|\rightarrow \infty$, indicating long
range order. 
A more interesting correlation function 
$C(r)=<\sin(\th({\bf r}_1))\sin(\th({\bf r}_2))>$
measures the fluctuations of the spins transverse to the
direction of the symmetry-breaking field.  We find
in all three phases
$C(r) \sim e^{-r/\la}$, but that in the log-confined (Log) phase,
$\la \equiv \la(\alpha)$, with $\alpha$ the angle between
${\bf r}$ and the symmetry axis of the underlying lattice.
(Throughout this study, we will assume the $XY$ spins reside
on a square lattice.)  This anisotropy in principle is a signature
that distinquishes the Log phase from the linearly confined
and deconfined phases.
We will see, however, that there is in addition an
anisotropy that exists in all three phases (and in the deconfined
phase of the $XY$ model without a symmetry-breaking field)
which renders the identification of the anisotropy due
to the vortex phase very challenging in numerical simulation.

The organization of this article is as follows. In the next section, 
we study correlation functions
in the three phases via some simple analytical models. 
Section III reports the results of our numerical simulations. 
Section IV discusses the anisotropy found
in our simulations for all the phases, which
we show is a ubiquitous effect of the
underlying lattice.  In Section V we report on
attempts to distinquish the lattice anisotropy from
that associated with the vortex phase in the numerical data.
Conclusions are presented in Section VI.

\section{Correlation functions: Analytical Models}
The Hamiltonian for the $XY$ model with a magnetic field is 
\be
H=-K \sum_{<{\bf r},{\bf r}'>} \cos \bigl[ \th({\bf r})-\th({\bf r}') \bigr]
- h \sum_r \cos \bigl[ \th({\bf r}) \bigr] ,
\ee
where  $<{\bf r},{\bf r}'>$ refers to nearest neighbor sites on a two dimensional square lattice.  
As discussed in the Introduction, we focus on
the order parameter and transverse fluctuation 
correlation functions $<\cos \bigl[ \th({\bf r}_1)\bigr ] \cos \bigl[\th({\bf r}_2)\bigr]>$ and
$<\sin \bigl[\th({\bf r}_1)\bigr] \sin \bigl[ \th({\bf r}_2) \bigr] >$.
Both these can be computed if we can 
calculate $<e^{i\th({\bf r}_1)\pm i\th({\bf r}_2)}>$.  A generating functional
that allows us to do this is
\bea
Z \bigl[ J \bigr]=<e^{i\sum_{\bf r} J({\bf r}) \th({\bf r})}>\\
\equiv \int  D\th e^{-H+i\sum_{\bf r} J({\bf r}) \th({\bf r})} / \int D\th e^{-H},
\eea
where $J({\bf r})=\de({\bf r})$ gives the order parameter average, and $J({\bf r})=
\de({\bf r}-{\bf r}_1)\pm \de({\bf r}-{\bf r}_2)$
generates $<e^{i\th({\bf r}_1)\pm i\th({\bf r}_2)}>$.

We may study our problem in the Villain model \cite{jose,villain,kadanoff}.
This essentially
involves replacing $e^{J\cos \theta}$ with 
$\sum_{m=-\infty}^{\infty} e^{-J(\theta -2\pi m)^2/2}$
wherever it appears in the partition function. 
The main idea is that the long distance physics depends only on the
symmetry of the Hamiltonian, not the
details of the interactions.


Following this prescription we may write
\bea
Z \bigl[ J \bigr]=\sum_{S_{rr'}}\sum_{T_r}\int \bigl[ D\th \bigr]
e^{-\f{1}{2K}\sum S^2_{rr'}-\f{1}{2h}\sum T^2_r}e^{i\sum S_{rr'}[\th(r)-\th(r')]+
i\sum_r(T(r)+J(r))\th(r)}\\
=\sum_{S_{rr'}}\sum_{T_r}\int \bigl[ D\th \bigr]
e^{-\f{1}{2K}\sum S^2_{rr'}-\f{1}{2h}\sum T^2_r}e^{i \sum (\na \cdot {\bf S} +T+J)\th(r)}.
\eea
Integrating out $\th$, we have
\be
Z \bigl[ J \bigr]
=\sum_{S_{rr'}}\sum_{T_r}
e^{-\f{1}{2K}\sum S^2_{rr'}-\f{1}{2h}\sum T^2_r} \de (\na \cdot S +T+J).
\ee
Note that the $\delta$-function appearing here is actually a Kronicker delta
with integer arguments.  The delta function allows a further simplification
if we write
\bea
S_\mu=\va_{\mu\nu}\pa_\nu n +{A}_\mu+\eta_\mu \\
\na \cdot {\bf A} {=T}\\
\na \cdot \eta=J ,
\eea
where $n$, $\bf{A}, \eta $ are integer fields. (Care must be taken in choosing
an allowed form of {\bf A} such that all configurations of the integer field 
{\bf S} are correctly produced \c{herb2}.)
From Eqs. (4) and (5),
we arrive at
\be
Z \bigl[ J \bigr]=\sum_{n,\bf{A}_\mu} e^{-\f{1}{2K}\sum |\va_{\mu\nu}\pa_\nu n+ {\bf A}_\mu +\eta_\mu|^2
-\f{1}{2h}\sum |\na \cdot {\bf A} |^2}.
\ee
 

\subsection{Linearly Confined Phase}
In the linearly confined phase, the integerness of $n$ and $A$ is unimportant \c{herb2},
and we may treat them as continuous fields. Integrating them out, we
find the generating functional takes the form
\be
Z[J]=e^{-\f{1}{2K}\int d^2 q \f{|J(q)|^2}{1/\xi^2+q^2} },
\ee
where $\xi=\sqrt{K/h}$. $\xi$ has the interpretation of the width of a string connecting
a vortex-anti-vortex pair \c{herb1,herb2}.

$\bullet$ Order parameter

For the order parameter, $J(r)=\de(r), |J(q)|=1$, and we get
\bea
<e^{i\th}>=\f{1}{[1+(\xi/a)]^{\pi/2K}}.
\eea

$\bullet$ Correlation function

For $<e^{i\th({\bf r}_1)-i\th({\bf r}_2)}>$, we take $J({\bf r})=
\de({\bf r}-{\bf r}_1)-\de({\bf r}-{\bf r}_2)$, and
using the fact that $<\sin(\th({\bf r}_1))\cos(\th({\bf r}_2)>=0$
by symmetry, we find
\be
<\cos(\th({\bf r}_1))\cos(\th({\bf r}_2))-\sin(\th({\bf r}_1))
\sin(\th({\bf r}_2))>
=e^{-\f{1}{2K}\int d^2 q \f{1-\cos(qr)}{1/\xi^2+q^2} },
\ee
where $r=|{\bf r}_1-{\bf r}_2|$.

Similarly, for $<e^{i\th({\bf r}_1)+i\th({\bf r}_2)}>$, 
we take $J({\bf r})=
\de({\bf r}-{\bf r}_1)+\de({\bf r}-{\bf r}_2)$, and find
\be
<\cos(\th({\bf r}_1))\cos(\th({\bf r}_2))+\sin(\th({\bf r}_1))
\sin(\th({\bf r}_2))>
=e^{-\f{1}{2K}\int d^2 q \f{1+\cos(qr)}{1/\xi^2+q^2} }.
\ee
Thus we have 
\be
<\cos(\th(r_1))\cos(\th(r_2))>
=\f{1}{2}[e^{-\f{1}{2K}\int d^2 q \f{1-\cos(qr)}{1/\xi^2+q^2} }+
e^{-\f{1}{2K}\int d^2 q \f{1+\cos(qr)}{1/\xi^2+q^2} }]
\ee
and 
\be
C(r)\equiv <\sin(\th(r_1))\sin(\th(r_2))>
=\f{1}{2}[e^{-\f{1}{2K}\int d^2 q \f{1-\cos(qr)}{1/\xi^2+q^2} }-
e^{-\f{1}{2K}\int d^2 q \f{1+\cos(qr)}{1/\xi^2+q^2} }].
\ee
For large distance r, we have approximately
\bea
<\cos(\th(r_1))\cos(\th(r_2))>
 \approx 1-\f{1}{2K}\int d^2 q \f{\cos(qr)}{1/\xi^2+q^2}\\
C(r)\equiv <\sin(\th(r_1))\sin(\th(r_2))>
\approx \f{1}{2K}\int d^2 q \f{\cos(qr)}{1/\xi^2+q^2} 
\eea
\be
C(r)\approx K_0(r/\xi)\sim \sqrt{\f{r}{\xi}}e^{-r/\xi},
\ee
where $K_0$ is the 0-th modified Bessel function, and the last form
of $C(r)$ is valid at large distances.

\subsection{Logarithmically Confined Phase}
In this phase, 
a suitable Hamiltonian for the long distance physics which
preserves the underlying lattice symmetry \c{herb2}
is 
\be
H=\f{1}{2K}\int d^2 q \{ [q_x^2+K\rho q_y^2]|\phi_1|^2+[q_y^2+K\rho q_x^2]|\phi_2|^2
+\xi^2 |q_x q_y(\phi_1+\phi_2)|^2 \},
\ee
where the $K \rho q_y^2 |\phi_1|^2, K\rho q_x^2 |\phi_2|^2$ 
terms are generated by the renormalization group flows of 
operators that 
model the 
integerness of the fields 
from which the continuous fields
$\phi_1,\phi_2$ were derived.
The generating function $<e^{i\int J \th}>$ can be calculated by the path integral
\be
Z[J]=\int D\phi_1 D\phi_2 \exp\{-\f{1}{2K}\int d^2 q \{ [q_x^2+K\rho q_y^2]|\phi_1|^2+[q_y^2+K\rho q_x^2]|\phi_2|^2
+\xi^2 |q_x q_y(\phi_1+\phi_2)+J|^2 \}\}.
\ee
Integrating out the fields $\phi_1,\phi_2$, we have
\be
Z \bigl[ J \bigr]=\exp\Biggl\{-\f{\xi^2}{2K}\int d^2q \f{|J(q)|^2}{1+\xi^2 q^2-
\xi^2 q^2\f{K\rho(q_x^4+q_y^4-q_x^2q_y^2)}{K\rho(q_x^4+q_y^4)+q_x^2q_y^2}}\Biggr\},
\ee
so that for large $r$,
\be
C({\bf r})\approx \f{\xi^2}{2K}\int d^2q \f{\cos(qr)}{1+\xi^2 q^2-
\xi^2 q^2\f{K\rho(q_x^4+q_y^4-q_x^2q_y^2)}{K\rho(q_x^4+q_y^4)+q_x^2q_y^2}}.
\ee
We see that in the Log phase, the mass terms $\propto K\rho$ generated
by the renormalization group flows introduce a characteristic
anisotropy in the correlation function $C({\bf r})$.  If evaluated at large
distance, $C({\bf r})$ depends exponentially with distance, but
the correlation length is angle-dependent due to the $K\rho$ term.

\subsection{Free vortex phase}
The free vortex phase occurs at high temperature, where the coefficients $K$ and $h$ are small.
The simplest way to understand the form of the correlation
functions is to return to the orginal $XY$ model and compute
the correlations functions perturbatively.
The transverse spin fluctuation correlation function is defined as
\be
C({\bf r}_1-{\bf r}_2)=<\sin(\th(r_1))\sin(\th(r_2))>=\int [D\th]\sin(\th_1)\sin(\th_2)
e^{K \sum_{<r,r'>} \cos \bigl[ \th(r)-\th(r') \bigr]
+ h \sum_r \cos \bigl[ \th(r) \bigr]}.
\ee
To second order in $K$ and $h$ we may expand this as
\bea
<\sin(\th({\bf r}_1))\sin(\th({\bf r}_2))>=\int [D\th]\sin(\th_1)\sin(\th_2) \\
\times \prod_{<{\bf r},{\bf r}^{\prime}>}
\{1+Kcos\bigl(\th({\bf r})-\th({\bf r}^{\prime})\bigr)
+\f{K^2}{2}[cos\bigl(\th({\bf r})-\th({\bf r}^{\prime})\bigr)]^2 \}
\prod_{\bf r}\{ 1+h\cos\th({\bf r})+\f{h^2}{2}\cos^2 \th({\bf r})\}\\
\approx \sum_{P_j} (\f{K}{2}+\f{Kh^2}{16})^{|P_j|},
\eea
where $P_j$ refers to self-avoiding paths connecting ${\bf r}_1$ and ${\bf r}_2$,
and $|P_j|$ is the length of the path. 
The shortest path gives the dominate contribution, so that 
$C(r)\sim e^{-r/\la}$, with correlation length
$\la=\f{1}{\ln(2/K)-h^2/8}$.
This same result can be obtained by approximating
the sum over self-avoiding paths as a sum over all random walks,
$\sum_{P_j} \rightarrow \sum_{P_j}^{\prime}$.
The summation over random walks can be calculated by a generating function
method \c{random}.  We first note that $C({\bf r})$ can be written as
\bea
C({\bf r})=\sum_{P_j}^{\prime}x^{|P_j|}\\
=\sum_N P_N ({\bf r})x^N
\eea
where $x=K/2+Kh^2/16$, ${\bf r}={\bf r_2}-{\bf r_1}=(m,n)$, $(m,n)$ are the coordinates of ${\bf r}$
and $P_N$ is the number of path of $N$ steps.  $P_N$ may be
conveniently written in terms of
the function $f(z_1,z_2)=z_1+z_1^{-1}+z_2+z_2^{-1}$, which obeys the identity
$(f)^N=\sum_{k,l} c^N_{k,l}z_1^k z_2^l$, where $P_N({\bf r})=P_N (m,n)=c^N_{m,n}$. Thus we
have
\be
P_N ({\bf r})=c^N_{m,n}=\f{1}{(2\pi i)^2}\int dz_1 dz_2 z_1^{m-1}z_2^{n-1}(f(z))^N.
\ee
Using eq.(21) and eq. (22) and setting $z_1=e^{iq_x}, z_2=e^{iq_y}$, we obtain 
\bea
C({\bf r})\approx \sum_N P_N x^N
=\int \f{dq_x dq_y}{(2\pi )^2} \f{e^{i {\bf q}\cdot{\bf r}}}{1-2x[\cos(q_x)+\cos(q_y)]}.
\eea  
For small $x$ and large $r$, the integral may be evaluated
leading again to the result
$C(r)\sim e^{-r/\la}$, with $\la=\f{1}{\ln(1/x)}=\f{1}{\ln(2/K)-h^2/8}$.

Using similar methods, the order parameter 
and order parameter correlation function may be
shown to behave as
\bea
<\cos(\th)>\sim \f{h}{2}, \\
<\cos\th(r_1)\cos\th(r_2)>\sim\f{h^2}{4}+e^{-r/\la},\\
\la=\f{1}{\ln(2/K)-3h^2/8}.
\eea

Finally, we note that similar results may be obtained within the
Villain model, although the derivation is considerably
more involved than this.

\section{numerical simulation}

In the previous section, we found a characteristic anisotropy in the Log
phase that is not present in either the linearly confined or unbound
vortex phases by computing correlation functions near 
fixed points of these phases \c{herb1,herb2}.  We now search 
for this behavior in simulations.  We shall see that 
anisotropy in the correlation functions is quite generic
in the $XY$ model, presumably due to irrelevant operators that were
not included in our effective Hamiltonians.  Our results suggest
that very high precision simulations are needed to detect the
effects of the vortex phase in the correlation functions.

We employ a standard Langevin dynamics approach for our simulations.
Angular variables $0\le \theta < 2\pi$ are represented on a
square lattice with periodic boundary conditions.  The model
is equivalent to that of an array of superconducting
grains, treated within the  
resistively shunted Josephson
junction model \c{sk}. 
The equations of
motion for the angular variables are
\be
\G\f{d^2 \th({\bf r})}{dt^2}=\f{\de H_{XY}}{\de\th({\bf r})}+\zeta({\bf r})-
\eta\f{d\th({\bf r})}{dt}.
\ee
The spin stiffness was taken as $K=1$, setting
our unit of energy.
The effective moment of inertial for the spins $\G$ were also taken to
be 1, setting the unit of time in our simulation. The viscosity
$\eta$  was taken to be 0.1, which we found yields good equilibration 
times and statistics.
$\zeta(\bf {r},t)$ is a random torque satisfying 
$<\zeta({\bf r},t)\zeta({\bf r}',t')>=2\eta
T\de_{{\bf r},{\bf r}'}\de(t-t')$ with $T$ being the temperature of the system (chosen
to be 1.2 for the results discussed below.)
The system size is $N_x=N_y=61$.
Our Hamiltonian $H_{XY}$
takes the form
 \be
H_{XY}=-K\sum_{<{\bf r},{\bf r}'>}cos[\th({\bf r})-\th({\bf r}')]
-h\sum_{\bf r} cos[\th({\bf r})],
 \ee
with $K=1$.
A typical run consists of $5\times 10^7$
time steps. Each time step is 0.1 (in units of $\sqrt{\G/K}$). 
We also eliminated the initial
$2\times 10^6$ steps in these runs for equilibration. We repeated runs for each set of
parameters with several different seeds, allowing us to estimate the
statistical error.

Our simulation results are summarized in Fig. 1-4. 
The curve with solid dots (squares) is for the transverse
spin fluctuation correlation function $C({\bf r})$ in the $x~(y)$
direction. Stars show correlation functions along a direction
with angle $\pi/4$ with respect to the $x$ axis. 
Fig. 1 shows results for $h$=0, while Figs. 2, 3 and 4 are respectively for 
$h=0.417,0.25$ and $0.15.$ These values were chosen to place the
system in the linearly confined, Log, and unbound vortex
phase, respectively, as found in an earlier study in
which the vortex phase was probed by measuring fluctuations
in the vortex dipole moment of the system \c{herb3}.

In all the simulations, we see clearly that 
$C(r) \sim e^{-r/\la}, r=|{\bf r}_1-{\bf r}_2|$;
i.e., the correlation function falls exponentially at large distances.
This agrees with our expectations from
Section II. The correlation functions in the $x$ and $y$ directions 
are the same within the numerical accuracy of the simulations,
as expected from the lattice symmetry.
There is also
a small difference between the amplitudes of the correlation functions 
along the bond direction and along the diagonal. However, the form of this
anisotropy is {\it not} what we expect from Section II, which suggested
anisotropy in the correlation length in the Log phase.  This should
lead to different slopes for $C(r)$ in the log-log plots along the
bonds and the diagonals at large distances.  Although there is some
evidence for this, it occurs only at very large distances, where our
statistics are relatively poor and an accurate measurement was not possible.

That most of the anisotropy appears as an offset in the correlation
functions in the log-log plots suggests that our results are dominated
by the prefactor of the exponential function for $C(r)$.  We explain
the origin of this in the next section, and argue that it is a generic
behavior that is independent of the vortex phase of the system.

\section{Lattice effects}

To take the lattice effects into account, the integration in eq. (7)
\be
\int d^2 q \f{|J(q)|^2}{1/\xi^2+q^2}
\ee
should be modified to
\be
\int d^2 q \f{|J(q)|^2}{1/\xi^2+4-2cos(q_x)-2cos(q_y)}
\ee
where our unit of length is the bond length.
Following the methods of Section II, the resulting
transverse spin fluctuation correlation function becomes
\be
C(r)\sim \int d^2 q \f{\cos qr}{1/\xi^2+4-2cos(q_x)-2cos(q_y)}.
\ee
For large distances, $C(r)$ may be calculated using
the stationary phase approximation \c{stat1,stat2}, with the result
\be
C(r)=\f{exp[-q_x^* m_x-q_y^* m_y]}{(|m|/R)^{1/2}} .
\ee 
Here ${\bf r}=(m_x,m_y)$, and $q_x^*, q_y^*$ are determined by 
\bea
1/\xi^2=4-2\cosh(q_x^*)-2\cosh(q_y^*)\\
m_y\sinh (q_x^*)=m_x \sinh (q_y^*),
\label{qstar}
\eea
and $R$ is given by
\be
R=\f{\sqrt{\sinh^2(q_x^*)+\sinh^2(q_y^*))}}{|\sinh^2(q_x^*)\cosh(q_y^*)+\sinh^2(q_y^*)\cosh(q_x^*)|}.
\ee
For large $\xi$, the solution to Eq. \ref{qstar} is
\bea
q_x=\f{1}{\xi}\cos\al\\
q_y=\f{1}{\xi}\sin\al,
\eea
where $\tan\al=m_y/m_x$. In this case, the correlation
length is $\la=\xi$ and 
\be
R=\f{\xi}{1+(a/\xi)^2\sin^2(\al)\cos^2(\al)}.
\ee
Thus we see that the lattice structure induces anisotropy in the prefactor
of the correlation function.
The ratio of the correlation function along two different directions,
with angles $\al_1, \al_2$ with respect to the $x$ axis, has the form 
\be
\f{C(r,\al_1)}{C(r,\al_2)}=\sqrt{\f{R(\al_1)}{R(\al_2)}}=1+\f{1}{8}(\f{a}{\xi})^2
(\sin^2(2\al_2)-\sin^2(2\al_1))
\ee
for $a/\xi<<1$.
This prefactor anisotropy
persists even at very long distance, and shows up as an offset in a plot
of $C(r)$ along different directions when displayed in a log-log
plot. This is exactly what we see in our numerical
simulation, and it is apparent that for the distances where we can
make reliable measurements, it is the dominent effect.
We note that, although this result was presented for
the linearly confined phase, one may show that in all the
phases discussed in Section II, a lattice regularization
of the momentum integrals leads to similar effects.

\section{Discussion}

As discussed above, there is anisotropy in the correlation functions for all
three phases due to lattice regularization. 
One can search for extra anisotropic
behavior in the Log phase, as in the section II.
To separate the two kinds of anisotropies,
we fit our numerical results by the formula
\be
C(r)=A \int d^2 q \f{\cos(qr)}{1+\xi^2 q^2-\f{K\rho q_x^2 q_y^2}{q^2}+aq^4
+bq_x^4q_y^4}.
\label{fit}
\ee
We choose the general form $\f{K\rho q_x^2 q_y^2}{q^2}$ (compatible with lattice
symmetry) to mimic the anisotropic behavior apart that should
not arise due to the lattice effects but is expected in the Log phase.
The term $aq^4$ is a higher order isotropic term and the term $bq_x^2q_y^2$ is the anisotropic
term due to lattice effects, expected from expanding the
cosines in a lattice regularization, and which for large
enough distances should be the leading order corrections
due to this regularization.
We fit our numerical curves for the correlation
function within a suitable distance interval
(generally from 3 to 6 correlation lengths.) 
If the distance is too small, short distance correlations 
become important and presumably need further terms to
be correctly modeled. On the other hand, when $r$ is too large, $C(r)$ 
becomes very small, and numerical errors spoil
the fit. The resulting fitting parameters
are shown in Table I.

\begin{table}
\caption{Fitting parameter I }

\begin{tabular*}{\textwidth}{c@{\extracolsep{\fill}}cccccc}\hline
$h$   & $A$  & $\xi$  & $K \rho$  & $a$  & $b$ \\ 
0.0 & 0.5165  & 4.6173  & -0.1940  & -0.8616  & 0.3821 \\
0.25 & 0.08154  & 1.7834  & -0.03805  & -0.1673  & 0.3765 \\
0.417 & 0.05268  & 1.4154  & -0.007307  & -0.1095  & 0.3060 \\ \hline

\end{tabular*}
\end{table}
We see that $K \rho$ is small but nonzero for all three phases, and
moreover is largest when $h=0$ when we expect no such term.
We  believe this is due to the fact that 
one needs to measure the correlation function
at much larger distances than we can achieve
numerically to see its true asymptotic behavior.
This is consistent with earlier simulations \c{herb3}
which required detection of vortex-antivortex
pairs at effective separations of order the system
size, typically $\sim$5 times larger than the maximum
$r$ used in our fits.
To support this point of view, we fitted our numerical 
results by Eq. \ref{fit}, but set $K \rho=0$ by hand; i.e.,
\be
C(r)=A \int d^2 q \f{\cos(qr)}{1+\xi^2 q^2+aq^4
+bq_x^4q_y^4}.
\label{fit2}
\ee 
The obtained fitting parameters are listed in Table II. 
 
\begin{table}
\caption{Fitting parameter II}

\begin{tabular*}{\textwidth}{c@{\extracolsep{\fill}}ccccc}\hline
$h$   & $A$  & $\xi$   & $a$  & $b$ \\ 
0.0 & 0.5163  & 4.6207  & -1.1499  & 3.0605 \\
0.25 & 0.08097  & 1.7884   & -0.1800  & 0.5610 \\
0.417 & 0.05256  & 1.4167  & -0.1110  & 0.3311 \\ \hline

\end{tabular*}
\end{table}

To further compare the fits with one another and the numerical data,
we use Eqs. \ref{fit} and \ref{fit2} to calculate model correlation functions
over long distances with the fit parameters. The difference between these 
turns out to be extremely small even at large distances.
(For $h=0.0$, it is $0.81\%$; for $h$=0.25, it is $4.9\%$; for $h$=0.417,
it is $4.0\%$.) The  absolute difference is $3.3\times 10^{-4}$ for $h=$0.0;
$2.8\times 10^{-5}$ for $h$=0.25; and $4.9\times 10^{-6}$ for $h$=0.417. 
Most importantly, in all cases these differences
are of the same order or smaller than our numerical error. 
To demonstrate this,
in Fig. 5 we compare the correlation functions $C(r)$ (both
along a bond direction and along the diagonal)  for $h$=0.25,
obtained by numerical simulation, from Eq. \ref{fit}, and from Eq. \ref{fit2}.
The curves from the three methods are nearly indistinquishable. In the
inset, we show the correlation function for an
arbitrary direction, $\tan( \al)=1/2$, where $\al$ is the angle 
with respect to the $\hat{x}$ axis.
Again, we can hardly see any difference among the curves.
Clearly, much higher precision data than we were able to
generate is required to distinquish the effects of lattice
anisotropy from the effects of logarithmically bound vortices.

\section{Conclusion}
We have discussed the effect of the various vortex phases
of the $XY$ model in a magnetic field on spin correlation
functions, focusing on transverse spin fluctuations.
Calculations based on the fixed point Hamiltonians
suggest that the Log phase should have a unique signature
in an anisotropic correlation length, but there is anisotropy
common to all the phases due the lattice realization of the
model, that shows up as irrelevant operators in the renormalization
group.  Distinquishing these lattice effects from those of the
vortex phase turns out to be quite difficult for system sizes
and run times that are readily attainable in our Langevin dynamics
simulations.  While we believe the anisotropy is present,
direct confirmation will require higher accuracies and larger
systems than were achieved in our own study.

{\bf Acknowledgments} 
This work was supported by the NSF 
through Grant Nos. DMR-0454699 and DMR-0511777.

\begin{figure}[b]
\includegraphics[width=15cm]{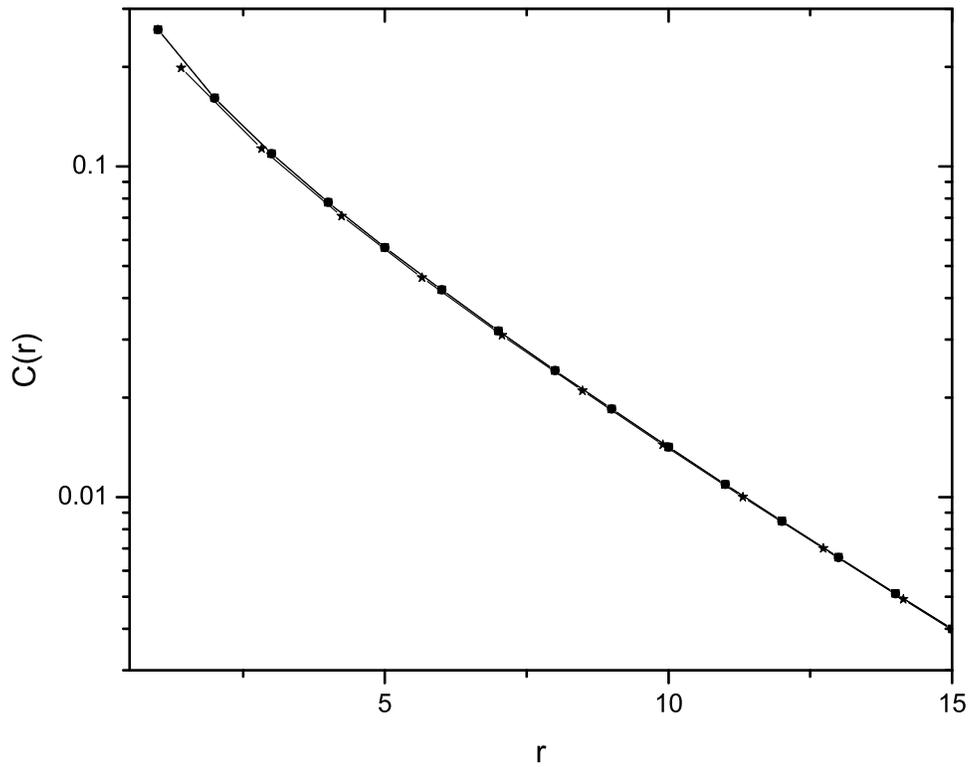}
\caption{Correlation functions vs. distance in a log-linear plot for $h$=0, system size=61. Averaged over 9 seeds. Curves with solid dots and squares are for correlation functions
in the $x$ and $y$ directions, and cannot be distinquised in the figure,
indicating the errors are smaller than the symbol sizes. Curve with stars is for
correlation function in the direction with angle $\pi/4$ with respec to $x$ axis.}
\end{figure}
~~~~~~~~~~~

\vfill\eject

~~~~~~~~

\begin{figure}[b]
\includegraphics[width=15cm]{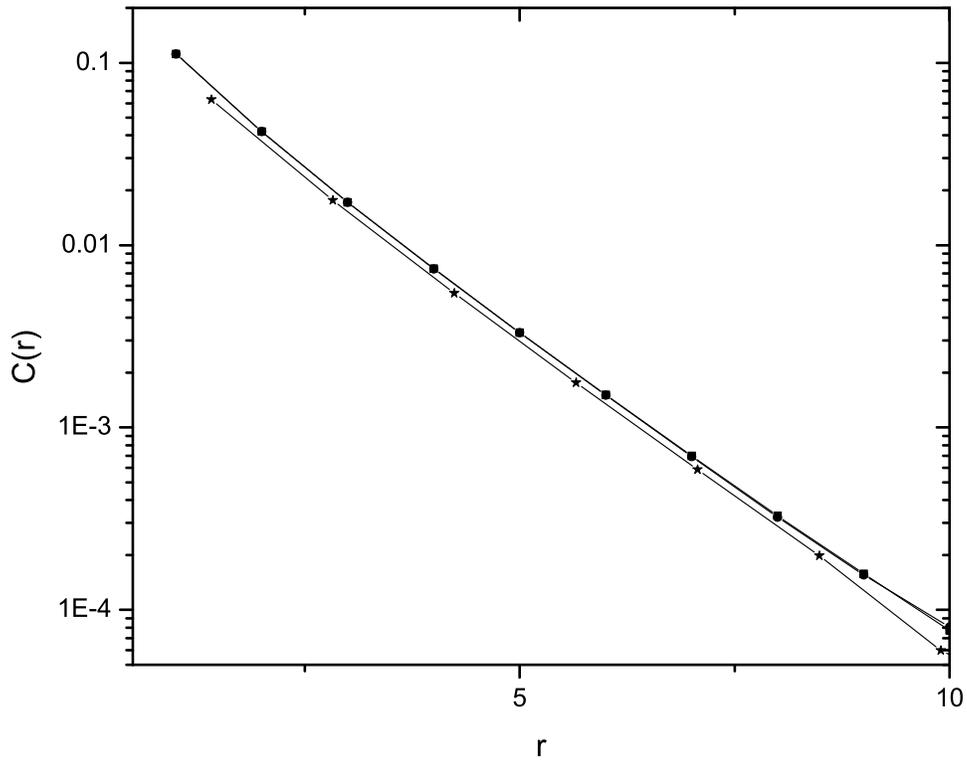}
\caption{Correlation functions vs. distance in a log-linear plot. $h$=0.417 (linearly confined
phase), system size=61. Averaged over 7 seeds. Curves with solid dots and squares are for correlation functions
in the $x$ and $y$ directions, and cannot be distinquised in the figure,
indicating the errors are smaller than the symbol sizes. Curve with stars is for
correlation function in the direction with angle $\pi/4$ with respec to $x$ axis.}
\end{figure}

~~~~~~~~~

\vfill\eject
\begin{figure}[b]
\includegraphics[width=15cm]{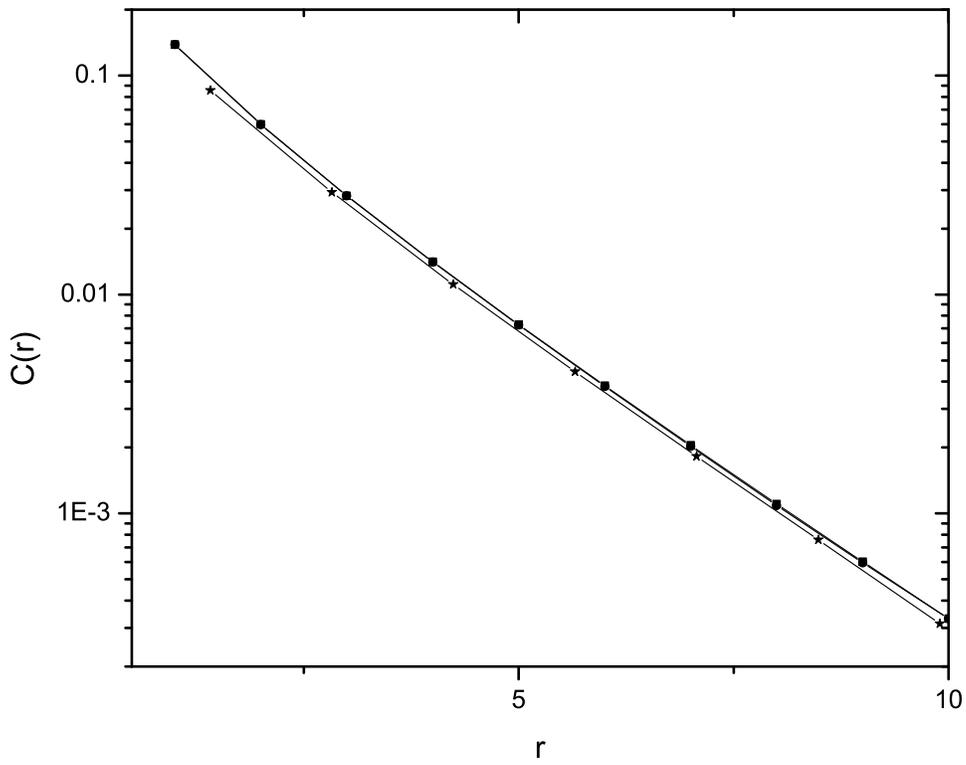}
\caption{Same as Fig. 2, except $h$=0.25 (Log confined phase), and data is averaged over 5 seeds.}
\end{figure}

~~~~~~~~~

\vfill\eject

\begin{figure}[b]
\includegraphics[width=15cm]{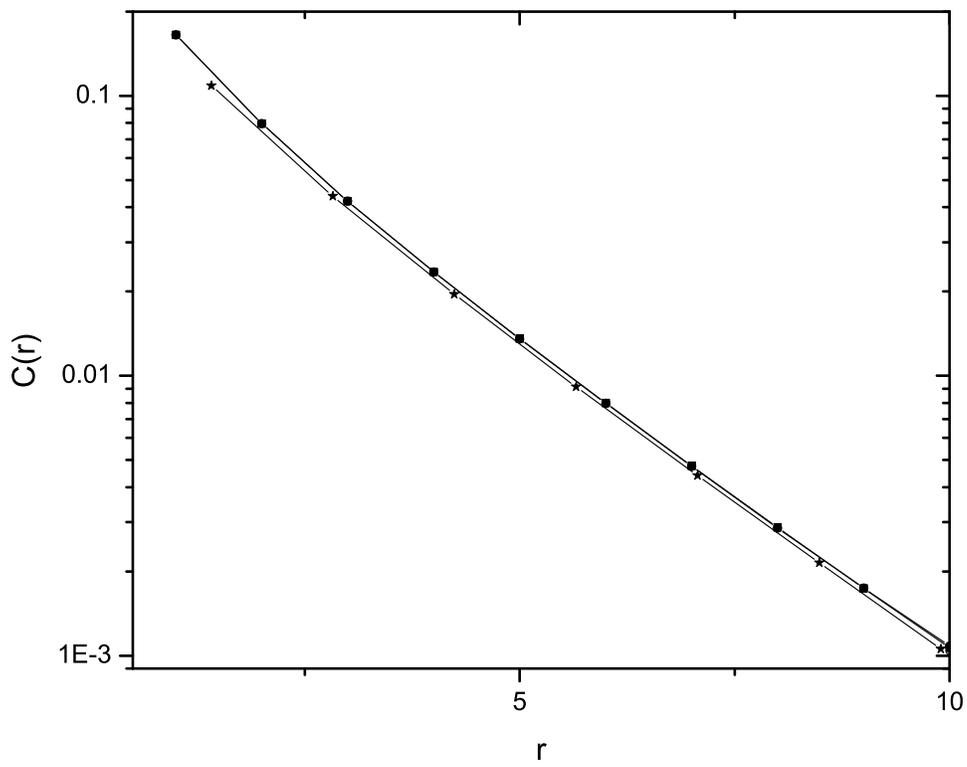}
\caption{Same as Fig. 2, except $h$=0.15 (free vortex phase), and data is
averaged over 2 seeds.}
\end{figure}

~~~~~~~~~

\vfill\eject

\vskip 5cm

\begin{figure}[b]
\vskip 5cm
\includegraphics[width=3cm]{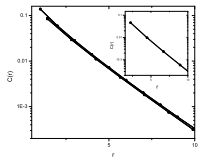}
\caption{Correlation functions vs. distance in a log-linear plot. $h=$0.25 (Log confined phase),
 system size=61. The two linesin the main panel (for $x$ direction and the direction with angle $\pi/4$ with respect to
$x$ axis are actually two pairs of three overlapping curves:
the results obtained from numerical 
simulation, fitting formula Eq. \ref{fit} and fitting formula Eq. \ref{fit2},
demonstrating that our data do not distinquish the expected anisotropy
in the Log phase from lattice anisotropy effects.
In the insert are 
correlation functions in the direction with angle $\tan \al=1/2$. Two curves obtained from Eqs. \ref{fit} and \ref{fit2}
are again indistinquishable.}
\end{figure}

\end{document}